\documentstyle[12pt,a4,epsf]{article}
\begin{document} 
\bibliographystyle{plain} 
\input{epsf} 
{\Large \bf \noindent Exact Results on Sinai's Diffusion } 
\vskip 2 truecm
\noindent Alain Comtet and \footnote{Current address: 
IRSAMC, Laboratoire de Physique Quantique, Universit\'e Paul Sabatier, 
 118 route de Narbonne, 31062 Toulouse Cedex}David S. Dean
\vskip 1 truecm
\noindent Division de Physique Th\'eorique, Institut de Physique
Nucl\'eaire, Universit\'e Paris-Sud, F-91406, Orsay Cedex, France.

\pagestyle{empty} 
\vskip 1 truecm \noindent{\bf Abstract:} 
We study the continuum version of Sinai's problem of a random walker in a random force field in one dimension. A method of stochastic representations is
used to represent various probability distributions in this problem (mean
probability density function and first passage time distributions). This method
reproduces already known rigorous results and also confirms 
directly some recent results derived using approximation schemes. We 
demonstrate clearly, in the Sinai scaling regime, that the disorder dominates
the problem and that the thermal distributions tend to zero-one laws.

\vskip 1 truecm
\noindent  August 1998
\newpage
\pagenumbering{arabic} 
\pagestyle{plain} 
\def\half{{1\over 2}}
\def\OO{\Omega}
\def\sech{{\rm sech}}
\def\n{{\newline}}
 \def\aa{\alpha}
 \def\bk{{\bf k}}
 \def\bkp{{\bf k'}}
 \def\bqp{{\bf q'}}
 \def\bq {{\bf q}}
 \def\EE{\Bbb E}
 \def\EEx{\Bbb E^x}
 \def\EEo{\Bbb E^0}
 \def\LL{\Lambda}
 \def\PP{\Bbb P^o}
 \def\rr{\rho}
 \def\SS{\Sigma}
 \def\ss{\sigma}
 \def\ll{\lambda}
 \def\dd{\delta}
 \def\ww{\omega}
 \def\ll{\lambda}
 \def\DD{\Delta}
 \def\DDt{\tilde {\Delta}}
 \def\kr{\kappa\lb \LL\rb}
 \def\PPx{\Bbb P^{x}}
 \def\gg{\gamma}
 \def\kk{\kappa}
 \def\tt{\theta}
 \def\bs{\hbox{{\bf s}}}
 \def\bh{\hbox{{\bf h}}}
 \def\lb{\left(}
 \def\rb{\right)}
 \def\prt{\tilde p}
\def\pt{\tilde {\phi}}
 \def\bb{\beta}
 \def\hal{{1\over 2}\nabla ^2}
 \def\bg{{\bf g}}
 \def\bx{{\bf x}}
 \def\bu{{\bf u}}
 \def\by{{\bf y}}
 \def\hag{{1\over 2}\nabla}
 \def\beq{\begin{equation}}
 \def\eeq{\end{equation}}
 \def\cosech{\hbox{cosech}}
\section{Introduction}
Since the introduction of the problem of a random walker in a random 
uncorrelated force field  by Sinai \cite{sinai} a great deal of 
activity has focused on this area. From the mathematical point
of view, in one dimension, many results may be obtained, for example
Kesten \cite{kesten} and Golosov \cite{gos} 
were the first to derive the average value
of the probability density function. Subsequently, results on the 
distribution of first passage times and consequently the probability
distribution of the maximum displacement were obtained. The model is one
of great interest as it is an analytically accessible test ground for the
study of non equilibrium phenomena in a non mean field situation where 
activation processes play a clear role. Recently a real space renormalization group study has been carried out which leads to results on multitime properties
of the process \cite{lefimo}, demonstrating explicity the phenomenon of aging in this model.
Physically the model finds applications in the study of the random field 
Ising model (which at low temperature may be viewed as a system of 
annihalating random walkers). The logarithmic diffusion seen in the Sinai 
model serves as a paradigm for the dynamics of domain walls in droplet models
for disordered spin systems \cite{fihu}.
A thorough discussion of the relevance of the 
Sinai model in condensed matter physics and a review of the early results
obtained may be found in \cite{boge,osetal}.

In this paper we analyse the continuum version of the Sinai model, using 
techniques similar to those used in the theory of disordered quantum systems
\cite{anpasl} and that have already been used in the study of the Sinai
model \cite{asetal,kata}.
We rederive the results already known in a very compact fashion which in 
addition sheds more light on the nature of these solutions. We derive
explicit representations for the distributions (and not just the disordered
averaged values). We shall see for example that the first passage time 
distributions become zero-one laws, i.e. the thermal fluctuations become 
insignificant and everything is dominated by the disorder.

The version of the 
model we shall study is that of a diffusion $X_t$ in a random force field.
The equation of motion is the Langevin equation

\beq 
{dX_t\over dt} = \eta_t + \phi'(X_t), \label{eq:sde0}
\eeq

where $\eta$ is a thermal white noise in time, i.e. it is Gaussian of zero mean
with correlation function $\langle \eta_t \eta_{t'}\rangle = 2T\delta(t-t')$.
The force $\phi'$ is derived from a potential $\phi(x)$ which is a 
Brownian motion of zero mean such that 
$\overline{\phi'(x)\phi'(y)} = \sigma^2 \delta(x-y)$. The result that the diffusion is logarithmic may be obtained through the following well known argument:
The time to move from the origin to a point $x$ is given by the Arrhenius law,
i.e.

\beq
t \sim \exp((\phi(x) - \phi(0)/T) \sim \exp(\sigma\sqrt{x}/T),
\eeq
by Brownian scaling, and hence $X(t) \sim \log^2(t)$. Exact results derived
show that this, rather crude, argument in fact reflects very accurately the
behaviour of the diffusion in the appropriate scaling regime. In this paper
the logarithmic nature of the diffusion is derived 
 from first principles without recourse to the
Arrhenius law or any scaling arguments.
\section{Mathematical Formalism}
We start with the model (\ref{eq:sde0}) 
of a particle moving in a quenched random white noise force
field $\phi'(x)$ such that $\overline{\phi'(x)\phi'(y)} = 2
\delta(x-y)$, here the overline indicates the average over the
disorder (i.e. without loss of generality we have taken $\sigma = \sqrt{2}$). 
The particle is also subject to a random white noise in time
such that the probability density function for a particle, started at
the origin at time zero, to be at $x$ at time $t$ is given by the
standard forward Fokker-Planck equation: 

\beq 
{\partial\over \partial
x}\lb {\partial p\over \partial x} + {d\phi\over dx}p\rb = {\partial
p\over \partial t}.  
\eeq 

In what follows we shall find it useful to
work with the Laplace transform of the density with respect to time
which is given by 

\beq 
{d\over dx}\lb {dp\over dx} + {d\phi\over dx}
p\rb = Ep -\delta(x), 
\eeq 

where $E$ is the Laplace transform variable
congugate to $t$. The adjoint $\rho(x)$ of the probability density $p(x)$ 
(the probability to be at $0$ given that the particle starts at
$x$) is related to $p(x)$ via detailed balance 

\beq p(x) = \rho(x)
\exp\lb -\phi(x) + \phi(0) \rb \label{eq:db} \eeq 
and can be shown to satisfy the backward equation

\beq {d^2 \rho\over dx^2} - {d\phi\over dx}{d\rho\over dx} =
E\rho - \delta(x).  
\eeq 

We now proceed by making essentially a
Hopf-Cole transformation of the problem

\begin{eqnarray}
\rho(x) &=& A \exp\lb -\int_0^x z_s^+ ds \rb \hbox{ for } x \geq 0 \nonumber\\
\rho(x) &=& A \exp\lb -\int_0^{-x} z_s^- ds \rb \hbox{ for } x \leq  0,
\end{eqnarray}

where it is easy to see that 

\beq {dz^\pm(s)\over ds} = -E + {z^\pm}^2(s) +
{d\phi(\pm s)\over ds}z^\pm(s) \ \ s\ge 0. \label{eq:sdez} \eeq 

The jump at $x=0$ fixes the
value of $A$ to be 

\beq A = {1\over z^+_0 + z^-_0}.  
\eeq 

Hence we
have found an explicit representation for $\rho$ (and hence $p$) in
terms of the two stochastic processes $z^\pm(s)$:

\begin{eqnarray}
\rho(x) &=&{1\over z^+_0 + z^-_0} \exp\lb -\int_0^x z_s^+ ds \rb
\hbox{ for } x \geq 0
\\ \rho(x) &=& {1\over z^+_0 + z^-_0} \exp\lb
-\int_0^{-x} z_s^- ds \rb \hbox{ for } x \leq 0
\end{eqnarray}

In general one requires boundary conditions for $z^\pm$, here we
imagine that the diffusion is restricted to a finite segment of the
real line $[-L,L]$ with reflecting boundaries at $x=\pm L$, hence 

\beq
{d\rho\over dx}\vert_{\pm L} = 0 \Rightarrow \; z^+_L = z^-_L = 0.
\eeq 

On using the equation of detailed balance (\ref{eq:db}) and the
integrated form of (\ref{eq:sdez}), one may show

\begin{eqnarray}
p(x) &=&{z_0^+\over z_x^+(z^+_0 + z^-_0)} \exp\lb -\int_0^x {E\over
z_s^+} ds \rb \hbox{ for } x \geq 0 \nonumber\\
 p(x) &=& {z_0^-\over z_{-x}^- (z^+_0 +
z^-_0)} \exp\lb -\int_0^{-x} {E\over z_s^-} ds \rb \hbox{ for } x \leq 0.
\end{eqnarray}

A useful property of $p$ is that we may write
 
\beq p(x) = -{1\over
E}{z_0^\pm\over(z^+_0 + z^-_0)}{d\over dx} \exp\lb -\int_0^{\pm x}
{E\over z_s^\pm} ds \rb, \eeq 

using this relation it is
straightforward to explicity verify that 

\beq 
\int_{-\infty}^{\infty}
p(x) dx = {1\over E}, 
\eeq 

i.e. the conservation of probability. The
observant reader will notice that the stochastic processes $z^\pm$
have the pathology that they have no normalisable equilibrium measure
and in addition that they may become infinite (finite time explosion).
Let us remark here that, for probabilistic reasons, $z$ must remain
positive as $\rho(x)$ must decrease monotonically on increasing the
distance from the origin. In fact the natural process to work with is
$w_s^\pm = z^\pm _{L- s}$,  which obeys the time reversed (with
respect to $z$) stochastic differential equation 

\beq 
{dw\over ds} = E
- w^2 + {d\phi\over ds}w \label{eq:sdew}.  
\eeq

Using the Stratonovich prescription
the generator for the diffusion $w$ is 

\beq 
{\cal G}_w =
{\partial\over \partial w}w {\partial\over \partial w}w - (w^2 -E)
{\partial\over \partial w}.  \eeq The equilibrium distribution for $w$
therefore satisfies \beq {\cal G}_w^{\dagger}P_0(w) = 0, 
\eeq 

and is
hence given by 

\beq 
P_0(w) = {1\over K_0(2E^{1\over 2})} {1\over
w}\exp(-w - {E\over w}),
\label{eq:eqw}
\eeq 
where $K_0(x)$ is a Bessel function of the third kind \cite{grry}.
It may be convenient, depending on the situation, to work with the
alternative variable 

\beq 
w = E^{1\over 2} \exp(\gamma), 
\eeq 

the fact
that we are using the Stratonovich prescription means that we may
apply the chain rule to the stochastic differential equation (\ref{eq:sdew})
and hence find: 

\beq {d\gamma \over ds} = -2E^{1\over 2}\sinh(\gamma) -
{d\phi\over ds} \eeq 

Slightly abusing notation, the equilibrium
distribution for $\gamma$ is given by: 

\beq P_0(\gamma) = {1\over
K_0(2E^{1\over 2})} \exp(-2E^{1\over 2}\cosh(\gamma))).
\eeq 

The
boundary conditions induced on $w$ are:
 
\beq 
{d\rho\over dx}\vert_{\pm L} =
0 \; \to \; z^\pm_L = w^\pm_0 = 0 
\eeq 

For example, for $x \geq 0$, $\rho$
is given in terms of $w$ as 

\beq \rho(x) = {1\over w^+_L +
w^-_L}\exp\lb - \int_{L-x}^L w^+_s ds\rb . 
 \eeq 

In the limit $L \to \infty$ the boundary conditions at $\pm L$ should become
unimportant. The following subtle argument demonstrates this fact.
On sending $L$ to infinity (whilst keeping $x$ finite)
the distribution of $w_L$  
is given by the equilibrium measure
(\ref{eq:eqw}) (i.e. $w_L$ is in equilibrium because $L$ corresponds to a large {\em time} by which the process $w$ is in equilibrium).
In addition, as $L\to \infty$ the distribution of $w_{L-x}$ is {\em also}
given by the equilibrium distribution.  
Hence treating $L-x$ as the new origin  for the process $w^+$, $L$ as a new
origin for $w^-$, and using the independence of $w^+$ and $w^-$
we may write the statistically identical expression:

\beq \rho(x) = {1\over w^+_x +
w^-_0}\exp\lb - \int_{0}^{x} w^+_s ds\rb,
 \eeq 

where it is to be
understood that $w^\pm_0$ are in equilibrium.

\section{Probability Density Functions}

With the formalism elaborated in the previous section it is now
immediate to calculate the probability density at the origin.
 This result has been calculated using the same
technique as here in \cite{asetal, kata} and we include it here for the sake of completeness as its derivation is now immediate. The same result may also be 
found in \cite{boetal} where it is calculated using the replica method from
disordered systems.

\beq \overline{\rho(0)} = 
\int_0^\infty {P_0(w) P_0(w')\over w + w'} dw
dw', \label{eq:intp0}
\eeq
and we start by expressing the denominator in the integral (\ref{eq:intp0}) as the
integral over an exponential, i.e.

\beq
\overline{\rho(0)} = {1\over E^{1\over 2}K_0(2E^{1\over 2})^2}
\int_0^\infty \exp\lb -t(x+y) - E^{1\over 2}(x + {1\over x})- E^{1\over 2}(y
+ {1\over y})\rb {dxdy\over xy} dt.
\eeq
The integrals over $x$ and $y$ may now be performed yielding (in terms of Bessel functions of the third kind \cite{grry})

\beq
\overline{\rho(0)} = {1\over E^{1\over 2}K_0(2E^{1\over 2})^2}
\int_{2E^{1\over 2}} ^\infty K_0^2(x)xdx.
\eeq
We now use the differential identity \cite{grry}
\beq 
{d\over dx}\lb {x^2\over 2}(K_0^2 - K_1^2)\rb = xK_0^2,
\eeq
thus yielding the result:

\beq
\overline{\rho(0)} = {4 K_1(2E^{1\over 2})^2 \over K_0(2E^{1\over 2})^2} -4
\label{eq:p0}.
\eeq
Using the asymptotic expansion for Bessel functions of the third kind 
\cite{grry} we find that for small $E$ (corresponding to large time) 

\beq
\overline{\rho(0)} \sim {1\over E \log^2(E)}, 
\eeq 

and hence using
the Tauberian theorems for regularly varying functions we obtain the result 

\beq
\overline{p(0,0,t)} \sim {1\over \log^2(t)}.  
\eeq 

To examine the full
$(x,t)$ dependence we note that 

\beq \overline{\rho(x)} = \int
{P_0(\gamma_0^+)P_0(\gamma_0^-)\over E^{1\over 2}\exp(\gamma^+) +
E^{1\over 2}\exp(\gamma_0^-)} G(\gamma_0^+, \gamma^+,x) d\gamma_0^+
d\gamma^+ d\gamma_0^-, \label{eq:ho1} 
\eeq 

where we have used the
Feynman-Kac Formula and $G$ is the Greens function obeying 

\beq
{\partial G\over \partial x} = {\partial^2 G\over \partial \gamma_0^2}
- 2E^{1\over 2} \sinh(\gamma_0){\partial G\over \partial \gamma_0} -
E^{1\over 2}\exp(\gamma_0) G, \label{eq:gf1} 
\eeq 

subject to the
initial condition 

\beq
 G(\gamma_0,\gamma, 0) = \delta(\gamma -
\gamma_0).  
\eeq 

In general the analysis of this Greens function is
difficult and few analytic results are known. However if we restrict
ourselves to the regime of Sinai scaling i.e we write $x= X\log^2(E)$ and
$\gamma = -u\log(E)$, where $u$ and $X$ are of order one, we find a
considerable simplification occurs which permits a rather thorough
asymptotic analysis of this regime; this is the key remark that enables us to
extend the method of \cite{asetal} to obtain new results. 
In these new variables, the
equation (\ref{eq:gf1}) becomes: 

\beq
 {\partial G\over \partial X} =
{\partial^2 G\over \partial u_0^2} + (E^{{1\over 2} - u_0} -
E^{{1\over 2} + u_0})\log(E){\partial G \over \partial u_0} -
\log^2(E) E^{{1\over 2} - u_0}G. 
\eeq

 In the region where $u_0 >
{1\over 2}$ we find the equation is dominated by the potential and drift terms
i.e.  

\beq 
{\partial G\over \partial X} = \log(E) E^{{1\over 2} -
u_0}{\partial G \over \partial u_0}   - \log^2(E) E^{{1\over 2} -
u_0}G,
\eeq 
which has the general solution
\beq
G(u_0,X) = {E^{u_0 - {1\over 2}}\over \log(E)}f(X + {E^{u_0 - {1\over 2}}\over
\log^2(E)}),
\eeq
and hence $G(u_0,X) \sim 0$ for $u_0 > {1\over 2}$. In the
region $u_0 <-{1\over 2}$, the equation is dominated by the drift term,
i.e.  

\beq {\partial G\over \partial X} = - E^{{1\over 2} +
u_0}\log(E){\partial G \over \partial u_0}, 
\eeq 

the general solution
to this equation being $G(u_0,X) = f(X+{1\over {E^{{1\over 2} + u_0}
\log^2(E)}})$ hence $G(u_0,X) \sim f(X)$, i.e. it becomes independent of
$u_0$. In the
interval $[-{1\over 2}, {1\over 2}]$ the equation is dominated by the
diffusive term, i.e.  

\beq 
{\partial G\over \partial X}= {\partial^2
G\over \partial u_0^2}.  
\eeq 

If we apply detailed balance in
(\ref{eq:ho1}) i.e. use 

\beq P_0(\gamma_0^+) G(\gamma_0^+, \gamma^+;x)
= P_0(\gamma^+) G(\gamma^+, \gamma^+_0;x) 
\eeq 

then we find 

\beq \overline{\rho(x)} = \int
{P_0(\gamma^+)P_0(\gamma_0^-)\over E^{1\over 2}\exp(\gamma^+) +
E^{1\over 2}\exp(\gamma_0^-)} F(\gamma^+,x)
d\gamma^+ d\gamma_0^-
\eeq 
where 
\beq 
F(\gamma^+,x) = \int_{-\infty}^{\infty} G(\gamma^+, \gamma^+_0,X)
d\gamma_0^+ .
\eeq 

By the Feynman-Kac formula $F(\gamma,x)$ obeys 

\beq 
{\partial F\over \partial x} =
{\partial^2 F\over \partial \gamma^2} - 2E^{1\over 2}
\sinh(\gamma){\partial F\over \partial \gamma} - E^{1\over
2}\exp(\gamma) F
\eeq 

with induced initial condition $F(\gamma,0) = 1$. Returning to the variable
$w$ we finally obtain:

\beq 
\overline{\rho(x)} = {1\over K_0(2E^{1\over
2})^2}\int_0^\infty {\exp(-w-w'-{E\over w} -{E\over w'}) \over ww'(w +
w')}F(\log({w\over E^{1\over 2}}),x)dwdw' .
\eeq 

We now exploit the fact
that $E$ is small

\begin{eqnarray}
\overline{\rho(x)} &=& {1\over E K_0(2E^{1\over 2})^2}\int_0^\infty
{\exp(-Ez-Ez'-{1\over z} -{1\over z'}) \over zz'(z + z')}F(\log({z
E^{1\over 2}}),x)dzdz' \nonumber \\ &\sim & {1\over E K_0(2E^{1\over
2})^2}\int_0^\infty {\exp(-{1\over z} -{1\over z'}) \over zz'(z +
z')}F(-\infty,x)dzdz' \nonumber \\ &=& {F(-\infty,x)\over E
K_0(2E^{1\over 2})^2}\int_0^\infty \exp(-u-u' - \alpha(u+u'))
dudu'd\alpha \nonumber\\ &=& {F(-\infty,x)\over E K_0(2E^{1\over
2})^2}. \label{eq:pf}
\end{eqnarray}

From our previous analysis we find that for $u \in [-{1\over 2}, {1\over 2}]$

\beq {\partial F\over \partial
X} = {\partial^2 F\over \partial u^2} . \label{eq:dif}
\eeq 

For $u_0 < -{1\over 2}$, $F$
is constant and continuity at $u = -{1\over 2}$ implies that
${\partial F\over\partial u}\vert_{u = -{1\over 2}} = 0$; hence $F(-\infty,x)
= F(-{1\over 2},x)$. Continuity
at $u = {1\over 2}$ yields $F({1\over 2},x) = 0$.  
The equation (\ref{eq:dif}) for $F$ in $[-{1\over 2}, {1\over
2}]$ may be easily solved in terms of Fourier modes to yield 

\beq
F(u,X) = {4\over \pi}\sum_{n=0}^\infty {1\over 2n +
1}\sin\lb({2n+1\over 2}) ({1\over 2} -u)\pi\rb \exp(-{(2n+1)^2\over 4}
\pi^2 X). 
 \eeq 

Note that if one had a Brownian motion in $[-{1\over
2}, {1\over 2}]$ started at a position $u$ with an absorbing boundary
at $u = {1\over 2}$ and a reflecting boundary at $u = -{1\over 2}$,
then $F(u,X)$ is simply the probability that the particle has not been
absorbed before time $X$, indeed the function $F$ appears directly in
the solution of Kesten \cite{kesten} 
with this very interpretation \cite{ledou}.

In the Laplace transformed variables one finds that 

\beq 
\overline
{p(x,E)} \sim {4\over \pi E\log^2(E)}\sum_{n=0}^\infty {(-1)^n\over 2n
+ 1} \exp(-{(2n+1)^2\over 4} \pi^2 {x\over \log^2(E)}).  
\eeq

Putting in the value of $F$ in equation (\ref{eq:pf}) and
asymptotically inverting the Laplace transform using the Tauberian
theorems we obtain the result of Kesten \cite{kesten}: 

\beq \overline {p(x,t)} \sim
{4\over \pi \log^2(t)}\sum_{n=0}^\infty {(-1)^n\over 2n + 1}
\exp(-{(2n+1)^2\over 4} \pi^2 {x\over \log^2(t)}) 
\eeq

\section{First Passage Time Distributions}

In this section we investigate the distribution of first passage times for
the Sinai diffusion, these results of course give information of the 
distribution of the maximum of the process.

Let us define $P(x,t)$ to be the probability starting at $x$ that the 
diffusion reaches the origin before time $t$. It is convenient to work with
$P$ evaluated at a random exponential time $T_E$ of rate E, i.e.

\beq P(x,E) = \int_0^\infty P(x,t) E\exp(-Et)dt = 
\langle \exp(-ET(x)) \rangle, \eeq

where the angled brackets indicate the thermal average (i.e. the average over the white noise $\eta$) and $T(x) 
= \inf\{ t\ :\ X_t =0 \ \vert \ X_0 = x\}$, i.e. the first passage time to  
$0$ starting from $x$.
A recurrence equation may now be found for $P(x,E)$ as follows, starting the 
process at $x$ evolve the particle for a time $dt$, then

\beq 
P(x,E) = \langle(1-Edt)P(x+dX_t,E) \rangle.
\eeq
Here the first term is simply the probability that the time $T_E$ does not occur during the first step and the second term is $P$ evaluated at the new position $x+ dX_t$. Expanding $P(x+dX_t,E)$ to order $dt$ and averaging over $dX_t$, one obtains the first passage time equation

\beq {d^2 P\over dx^2} - {d\phi\over dx}{dP\over dx} =
EP   
\eeq 

Using the boundary condition $P(0,E) = 1$ and making the Hopf-Cole transformation of the previous section we find:

\beq 
P(x,E) = \exp(-\int_0^x z_s ds).
\eeq

where $z_s$ obeys the stochastic differential equation (\ref{eq:sdez}). 
Re-expressing the above in terms of the process $w$ we find

\beq 
P(x,E) = \exp(-\int_{L-x}^L w_s ds).
\eeq

Consider $x=L$, i.e. we consider the time taken to cross the sample with reflecting initial conditions at the starting point (hence $z_L = w_0 = 0$).
If we consider average value of the $q^{th}$ power of $P(L,E)$ i.e. 
$\overline{P^q(L,E)}$ 
for fixed
$L$ and $E$, using the Feynman-Kac formula one finds that   
$\overline{P^q(L,E)} = F_q(L,E)$ where
\beq 
{\partial F_q\over \partial x} =
{\partial^2 F_q\over \partial \gamma_0^2} - 2E^{1\over 2}
\sinh(\gamma_0){\partial F_q\over \partial \gamma_0} - qE^{1\over
2}\exp(\gamma_0) F_q.  
\eeq 
 
It is now important to note that for $q >0$, in the Sinai scaling regime, the prefactor $q$ does not alter the solution for $F_q$, hence one has that
$\overline{P^q(L,E)} = \overline{P(L,E)}$, that is to say that
$P$ tends to either zero or one  independently of the thermal noise i.e.
the distribution of $P(L,E)$ for fixed $L$ and $E$, 
${\cal{P}}$ tends to the form ${\cal{P}}(P) =
\alpha \delta(P-1) + (1-\alpha)\delta(P)$. Using the calculation of the 
previous section for $F$ we find therefore, in the regime of large times and 
in the Sinai scaling regime where $L/\log^2(E) \sim O(1)$, that 

\beq 
\alpha =  {4\over \pi}\sum_{n=0}^\infty {(-1)^n\over 2n
+ 1} \exp(-{(2n+1)^2\over 4} \pi^2 {L\over \log^2(E)}).
\eeq

One may also consider the case where $L\to \infty$, in this limit we see
that 
\beq 
P(x,E) = \exp(-\int_{0}^x w_s ds). \label{eq:pe}
\eeq
but where the boundary condition on $w_0$ is that it is in equilibrium with the measure (\ref{eq:eqw}), here we therefore find
\beq 
\overline{P^q(x,E)} = \overline{P(x,E)} = \int_{-\infty}^\infty P_0(\gamma_0)
F(\gamma_0, {x\over \log^2(E)}).
\eeq
In the Sinai scaling regime $x/\log^2(E) \sim O(1)$ we may therefore write:
\beq \overline{P(x,E)} = 
{1\over K_0(2 E^{1\over2})}\int_{-\infty}^\infty \exp( -E^{{1\over 2} - u_0} - E^{{1\over 2} + u_0} ) F(u_0) \vert \log(E)\vert du_0,
\eeq
which becomes, in the limit of small $E$,
\beq
\overline{P(x,E)} = \int_{-{1\over 2}}^{1\over 2} F(u_0) du_0 = 
{8\over \pi^2}\sum_{n=0}^\infty {1\over (2n+1)^2}\exp\lb -{(2n+1)^2\over 4}
{\pi^2 x\over \log^2(E)}\rb .
\eeq
Here it is interesting to note that one may calculate explicitly 
$\overline{\log(P(x,E))}$ which is given as
\beq
\overline{\log(P(x,E))} = - x\int_0^\infty P_0(w)wdw \ ,
\eeq
which in the limit of small $E$ (or large time), but not dependent on being in the Sinai scaling regime, yields
\beq
\overline{\log(P(x,E))} \sim - {x\over \log(E)},
\eeq
hence, whereas the scaling of the integer moment of $P(x,E)$ was as
${x\over \log^2(E)}$, the scaling of the average of the logarithm is as
${x\over \log(E)}$, suggesting that in a typical sample the scaling should be
as ${x\over \log(E)}$. If we consider the case where $x$ is infinitesimally
close to the origin we see that the probability to return to the origin
(simply developing the exponential in equation (\ref{eq:pe}) to first order
in $x$) is  
\beq
\overline{P(x,E)} = 1 - {x\over \log(E)},
\eeq
and hence following the definition of \cite{lefimo,igri} the probablity 
of not arriving at the 
origin behaves as $p_{pr}(L) \sim L^{-\theta}$ where $L$ is the length scale
$L= \log^2(E)$ and $\theta$ is therefore the persistence exponent for the 
diffusion and is given by $\theta = {1\over 2}$, thus confirming the results 
of \cite{lefimo, igri}.  
\section{The Mean First Passage Time}
In the Sinai model we recall that there are two kinds of average, a thermal 
average over the realisations of the temporal white noise and also the average
over the realisations of the disorder. There has been much interest in the
distribution of the mean first passage time (MFPT)
. Various authors have remarked upon the multifracticality of its distribution, in this section we present an 
extremely simple argument to demonstrate this fact. Consider the case where
there is a reflecting boundary at $L$, using the formalism of the 
previous sections or using the formula for MFPTs in \cite{gar} we have

\beq 
{d^2\over dx^2}\langle T_L(x) \rangle - {d\phi\over dx}{d\over dx}\langle T_L(x) \rangle = -1,
\eeq

where the subscript $L$ indicates the reflecting boundary conditions at $L$
and $T_L(x)$ is the first time the particle arrives at $0$ given that it
starts at $x$ (the angled brackets indicating the thermal average).
The solution to this equation for a particle starting at $L$ is

\beq
\langle T_L(L) \rangle = \int_0^L dx \exp(\phi(x))\int_x^L dy \exp(-\phi(y)).
\eeq

The moments averaged over the disorder are 

\begin{eqnarray}
\overline{\langle T_L(L) \rangle^q} &=& 
\int_{\{ x_n < y_n\}} \prod_1^q dx_n dy_n \exp\lb \sum_{ij} (\min(x_i,x_j) -
\min(y_i,y_j) - 2\min(x_i,y_j))\rb \nonumber \\
&=& \int_{\{ x_n < y_n\}} \prod_1^q dx_n dy_n \exp\lb -{1\over 2}
\sum_{ij} (\vert x_i-x_j \vert +  \vert y_i-y_j \vert - 2
\vert x_i-y_j \vert)\rb.
\end{eqnarray}
Hence we see that the moment $\overline{\langle T_L(x) \rangle^q}$ is 
the partition function for a one dimensional anti-Coulomb gas (in the
sense that like charges attract and opposite charges repel), where
each particle has its antiparticle which is conditioned to be on its right.
Rescaling the spatial variables we obtain

\beq 
\overline{\langle T_L(L) \rangle^q} = L^{2n} 
\int_{\{ x_n < y_n\}} \prod_1^q dx_n dy_n \exp\lb -{L\over 2}
\sum_{ij} (\vert x_i-x_j \vert +  \vert y_i-y_j \vert - 2
\vert x_i-y_j \vert)\rb,
\eeq
where all the integrations are now in $[0,1]$. When $L$ becomes large 
this corresponds to the limit of low temperature and the partition function
is dominated by the lowest energy state which is clearly when $x_n =0$ and 
$y_n = 1$ for all $n$, i.e. all the like charges condense together to form
two macrocharges which repel and hence lie on the two boundary points
of the unit interval. This energy is simply $q^2 L$. Hence

\beq 
\log \lb \overline{\langle T_L(L) \rangle^q} \rb \sim q^2 L,
\eeq
clearly demonstrating the multifracticality. The distribution is 
infact approximately log-normal.

\section{Probabilistic Interpretation of Results}
Here we shall interpret the results derived in this paper in a probabilistic
fashion which gives us some intuition for the structure of the Sinai problem.
In terms of stochastic processes we shall consider the process $u_x
= -\gamma_x/\log(E)$; if we define the rescaled time variable $x = X\log^2(E)$
then the stochastic equation obeyed by $u$ is

\beq
{du\over dx} = {1\over\log^2(E)} {du\over dX} = -{1\over \log(E)}\lb E^{{1\over 2}+u}
- E^{{1\over 2}-u} - {1\over \log^2(E)}{d\over dX}\phi(X\log^2(E)) \rb.
\eeq
Using Brownian scaling we may write $\phi({X\log^2(E)}) \equiv \phi(X)
 \vert\log(E)\vert$ and hence the induced stochastic differential equation on
$u$ as a function of the rescaled time $X$ is

\beq
{du\over dX} = -\log(E)(E^{{1\over 2}+u}
- E^{{1\over 2}-u}) + {d\over dX}\phi(X)
\eeq

In the region outside the interval $[-{1\over 2}, {1\over 2}]$ the process
is dominated by a drift which returns the particle to the interval. In the 
interval $[-{1\over 2}, {1\over 2}]$ the process is simply a Brownian motion.
If we now consider the functional critical to our study of this problem
\beq
H(X) = \exp\lb -\int_0^x w_s ds \rb \equiv \exp\lb -\log^2(E)\int_0^X E^{{1\over 2} - u_s} ds\rb,
\eeq
we see that as $E\to 0$
\begin{eqnarray}
H(X) &\to & 1 \ \ \hbox{if} \ \max(u_s) < {1\over 2} \nonumber \\
     &\to & 0 \ \ \hbox{if} \ \max(u_s) > {1\over 2},
\end{eqnarray}
thus clarifying at an intuitive level the zero-one type results obtained 
in the previous sections.

\section{Conclusions and Perspectives}
We have shown that the method presented in \cite{asetal,kata} 
(and similar to those used in the Quantum Mechanics of one dimensional disordered systems \cite{anpasl}) may be extended to 
produce explicit results on spatio-temporal probability distribution functions in the continuum version of the Sinai model; notably the result of Kesten 
for the averaged probability density function and also for various first passage time probabilities. The method demonstrates the preponderence of zero-one type laws where the disorder dominates. The continuum formulation of the problem also allows
a very simple demonstration of the multi-fractal nature of the mean first 
passage time. 

The great advantage of the technique is that it allows one to represent 
transition probabilty densities as functionals of a well defined stochastic
process and therefore quantities depending on several times may in principle 
be analysed, thus permitting the study of aging in these systems. Work on
these extensions is currently under progress \cite{code}.

\pagestyle{empty}

\end{document}